\documentstyle[graphicx]{mn}
\begin{document}
\title{Fast magnetosonic waves in pulsar winds}
\author[Y.E.Lyubarsky]{Y.E.Lyubarsky\\
Physics Department, Ben-Gurion University, P.O.B.\ 653, Beer-Sheva
84105, Israel; e-mail: lyub@bgumail.bgu.ac.il}
\date{Received/Accepted}
\maketitle
\begin{abstract}
 Fast magnetosonic waves in a
magnetically-dominated plasma are investigated. In the pulsar
wind, these waves may transport a significant fraction of the
energy flux. It is shown that the nonlinear steepening and
subsequent formation of multiple shocks is a viable mechanism for
the wave dissipation in the pulsar wind. The wave dissipation both
in the free pulsar wind and beyond the wind termination shock is
considered.
\end{abstract}


\section{Introduction}
The mechanisms of the energy transfer from the pulsar to the
pulsar wind nebula still remain obscure. It is widely accepted
that pulsars emit an electron-positron plasma, which form an
ultrarelativistic magnetized wind. The rotational energy of the
neutron star is carried mostly by electromagnetic fields as
Poynting flux (Michel 1982) and should be eventually transferred
to the radiating particles. The wind terminates at a shock front
located, in the case of the Crab Nebula, some $10^{17}$ cm from
the pulsar. The postshock flow parameters may be matched with the
observed Crab structure if the energy flux at the shock front is
carried mainly by the particles (Rees \& Gunn 1974; Kennel \&
Coroniti 1984; Emmering \& Chevalier 1987). {\bf However it is
extremely difficult for the pulsar wind to reach the kinetic
energy dominated state, at least if the wind is considered as an
ideal MHD flow} (Tomimatsu 1994; Bogovalov 1997; 2001b; Beskin,
Kuznetsova \& Rafikov 1998; Chiueh, Li \& Begelman 1998; Lyubarsky
\& Eichler 2001). The necessary conversion may be provided by
dissipation mechanisms. These mechanisms operate faster at small
scales and first affect waves generated in the wind by the
rotating, oblique pulsar magnetosphere.

There is a variety of electromagnetic waves in rarefied magnetized
plasmas (e.g., Akhiezer et al.\ 1975) however we assume that only
true MHD waves (those satisfying the condition $\bf E=v\times B$)
may be generated by the rotating magnetosphere. The reason is that
according to the convenient view, the plasma density in the pulsar
wind is sufficiently large such that {\bf low-frequency}
electromagnetic waves are heavily damped (e.g., Asseo et al.\
1978; Melatos \& Melrose 1996). There are four types of MHD waves
but only the entropy and the fast magnetosonic (FMS) waves may
propagate far beyond the light cylinder where the magnetic field
is predominantly toroidal. {\bf In an oblique magnetosphere, the
magnetic field at the light cylinder oscillates with the pulsar
rotation period. These oscillations propagate outwards as MHD
waves. In the equatorial belt of the flow, the magnetic field line
at a given radius alternates in direction with the frequency of
rotation, being connected to a different magnetic pole every
half-period. Such an alternating field is transported by the flow
in the form of the striped wind (Michel 1971, 1982; Bogovalov
1999), which may be considered as an entropy wave. At high
latitudes, where the magnetic field does not change sign, the
magnetic oscillations are transferred away in the form of FMS
waves (generation of FMS waves by the rotating, slightly
nonaxisymmetric magnetosphere was considered by Bogovalov 2001a)}.

The entropy wave decays because of the current starvation in
current sheets separating strips with the opposite magnetic field
(Usov 1975; Michel 1982, 1994; Coroniti 1990). Lyubarsky \& Kirk
(2001) showed that the flow significantly accelerates in the
course of reconnection and this dilates the timescale over which
the wave decays. At typical conditions, the dissipation radius
exceeds the radius of the termination shock therefore one should
conclude that the Poynting flux in the striped wind does not
dissipate until the wind enters the termination shock. All the
energy should release within the shock where the flow decelerates.

In this article, the fate of FMS waves is considered. It will be
shown that these waves may be described within the MHD framework
throughout the pulsar wind up to the termination shock and ever
beyond. The reason is that in a magnetically dominated plasma, the
FMS waves excite small conductivity currents, the oscillations of
the magnetic field being nearly compensated by the displacement
current. In MHD regime, FMS waves may decay due to the nonlinear
steepening and subsequent formation of multiple shocks. It will be
shown that for typical pulsar parameters, the waves may decay only
beyond the termination shock. However in rapidly spinning pulsars,
like the Crab, FMS waves may decay before the flow reaches the
termination shock provided the plasma density in the wind is high
enough.

The article is organized as follows. Properties of FMS waves in a
magnetically dominated plasma are outlined qualitatively in Sect.\
2. These estimates are applied to the pulsar wind in Sect.\ 3. The
wave decay after the multiple shocks formation is considered in
Sect.\ 4. Conclusions are summarized in Sect.\ 5. In Appendix 1,
exact solutions for nonlinear FMS waves in the magnetically
dominated plasma are presented. In Appendix 2, the energy and
momentum fluxes transferred by FMS waves are derived.

\section{FMS waves in a magnetically dominated plasma}
 In this section, I consider FMS waves in the wind frame. Let us
consider waves propagating perpendicularly to the magnetic field
because far enough from the pulsar, the magnetic field is nearly
toroidal whereas the waves propagate radially. Only qualitative
estimates are outlined here; the exact solutions are presented in
Appendix 1.

When the FMS wave propagates perpendicularly to the magnetic
field, plasma oscillates along the wave direction together with
the magnetic field. The flux freezing condition ties the plasma
density and the magnetic field together:
$$
\frac{B'}{n}=\frac{B'_*}{n_*}\equiv b=\it const. \eqno(1)
$$
Here $B'$ and $n$ are the magnetic field and the plasma density in
the proper plasma frame, {\bf the subscript * stands for
quantities averaged over the wave period. The quantities measured
in the proper plasma frame differ from those measured in the wind
frame because the plasma oscillates. In this section, I neglect
this difference assuming the oscillation velocity to be
non-relativistic. More general consideration is presented in
Appendix 1.} With the aid of Eq.(1), one can express the magnetic
energy and the pressure via the plasma density thus considering
the wave as the sound wave in a medium with the equation of state
$$
{\cal P}=p+\frac{b^2n^2}{8\pi};\qquad {\cal
E}=\varepsilon+\frac{b^2n^2}{8\pi},
$$
where $p$ and $\varepsilon$ are the plasma pressure and energy
density, correspondingly. The velocity of the wave may be now
found immediately as the sound velocity in such a
medium(throughout the paper all velocities are expressed in units
of the speed of light, $c=1$) :
$$
s_M^2=\frac{d\cal P}{d\cal E}
=\frac{ws^2+b^2n^2/4\pi}{w+b^2n^2/4\pi}, \eqno(2)
$$
where $w=p+\varepsilon$ is the enthalpy of the plasma,
$s=\sqrt{dp/d\varepsilon}$ the thermal sound velocity.

In the pulsar wind, the magnetic energy density significantly
exceeds the plasma energy density. One can conveniently express
all the values via the parameter
$$
\sigma=\frac{b^2n^2}{4\pi w}, \eqno(3)
$$
which is twice the ratio of the magnetic to the plasma energy
density in the wind frame and the ratio of the Poynting flux to
the plasma energy flux in the laboratory frame. One can see that
at $\sigma\gg 1$ the FMS velocity is close to the speed of light.
The corresponding Lorentz factor, $\gamma_M\equiv
(1-s_M^2)^{-1/2}$, may be written as
$$
\gamma_M=\sqrt{\frac{\sigma}{1-s^2}}.
$$

In the small amplitude, linear wave, the electric field may be
written as $E'=vB'_*$ whereas variations in the magnetic field are
$\delta B'=B'_*(\delta n/n_*)$. Taking into account that
variations in the density and in the velocity are related by the
standard expression $\delta n/n_*=v/s_M$, one can see that
$$
E'=\frac{\delta n}{n_*}=\delta B'\eqno(4)
$$
to within a factor of $1/\sigma$. Because the electric field may
not exceed the magnetic field, the wave amplitude should be
limited by
$$
\delta B'<B'_*/2. \eqno(5)
$$
It is shown in the Appendix 1 that when the wave amplitude
approaches the limiting value, the oscillation velocity becomes
ultrarelativistic, the plasma density goes to zero at some phase
of the wave period and moreover the characteristic scale for the
nonlinear steepening of the wave decreases to zero.

Now let us find what plasma density is necessary to allow the MHD
solution. The plasma density should satisfy the evident condition
$en>j$, where $j$ is the conductivity current excited in the wave.
Substituting into the Maxwell equations
$$
{\bf\nabla\times B'}=4\pi{\bf j}+\frac{\partial\bf E'}{\partial
t}; \qquad {\bf\nabla\times E'}=-\frac{\partial\bf B'}{\partial
t},
$$
a sinusoidal wave $\exp(-i\omega t+{\bf k\cdot x})$ with the
dispersion law $\omega/k=s_M$, one can find the conductivity
current in this wave,
$$
j=(1-s_M^2)i\omega E'/4\pi.
$$
One can see that in high-$\sigma$ flows, the conductivity current
in FMS waves is $\sigma$ times less than the displacement current.
Substituting the maximal current density, $j=en$, and making use
of Eqs.(2--4), one can see that the cold plasma may provide the
necessary conductivity current if
$$
\omega_B>\omega\frac{\delta B'}{B'_*}, \eqno(6)
$$
where $\omega_B=eB'_*/m$ is the  Larmour frequency. We consider
the electron-positron plasma therefore throughout the paper $m$ is
the electron mass.

%Note that violation of the condition (5) does not result in violent particle
%acceleration provided the condition (4) still holds. Particles do accelerate in the
%strong vacuum electromagnetic wave because $E=B$ in this wave. If the vacuum wave
%propagates in the background magnetic field such that the condition (4) holds
%(the wave magnetic field is assumed to be directed along the background field),
%particles experience only drift oscillations with nonrelativistic velocities.

At the conditions (5, 6),  FMS waves in a magnetically dominated
plasma may be considered in MHD approximation. In the MHD regime,
the wave may decay as a result of nonlinear steepening and
subsequent formation of multiple shocks. The wave steepening
occurs because the wave velocity depends on the plasma density and
therefore the compressive part of the wave moves faster than the
expansive one. In the strongly magnetized plasma, even a large
amplitude wave is only weakly nonlinear because the wave velocity
varies with the density only by a factor of about $1/\sigma$. The
reason is that, as it was shown above, the conductivity current,
which only introduces nonlinearity into the MHD equations, is
small in the magnetically dominated plasma.

The difference in the velocities between the compressive and
expansive parts of the wave may be estimated from Eq.(2) as
$$
\delta s_M\sim\frac 1{\sigma}\frac{\delta n}{n_*}.
$$
The shock forms when the leading edge of the wavefront becomes
vertical (e.g., Landau \& Lifshitz 1959). This occurs within a
characteristic nonlinear time it takes from the compressive part
to shift relative the expansive one by $\sim 1/\omega$. Therefore
the shock forms after the wave travels a distance
$$
x_{\rm nl}\sim 1/(\delta s_M\omega)\sim\sigma/\omega. \eqno(7)
$$
This estimate is confirmed by a rigorous derivation in Appendix 1.
After the shock formation, the wave continue to distort at the
characteristic scale $x_{\rm nl}$ such that eventually all the
wave energy dissipates in the shock. However both $\sigma$ and
$\omega$ may change because the plasma heats up and accelerates
(or decelerates) considerably in the course of the wave
dissipation. Therefore numerically the dissipation scale may
differ from the shock formation scale even though the same
expression (7) is valid for both scales.

\section{FMS waves in the pulsar wind}
Far beyond the light cylinder, the wind may be considered as
purely radial, whereas the magnetic field as purely toroidal. The
wind is assumed to be super-FMS, $\gamma>\sqrt{\sigma}$.
Transforming the electromagnetic fields from the wind frame into
the pulsar frame, one can see that the condition (5) reduces to
the condition $\delta B<B_*$ (the factor of two appears because in
the proper plasma frame $E'_*=0$ whereas in the pulsar frame
$E_*=v_*B_*\approx B_*$). So FMS waves may be generated by the
rotating oblique magnetosphere at high latitudes where the
magnetic field does not change the sign. In the equatorial belt
(its width depends on the angle between the magnetic and the
rotation axes of the pulsar) an entropy wave is generated in the
form of the striped wind. Generally a superposition of an entropy
and FMS waves is generated here (one can talk about the
superposition because even a large amplitude FMS wave is weakly
nonlinear in the magnetically dominated plasma) however an
important point is that an entropy wave arises here inevitably
because there is no other MHD wave that may transfer alternating
magnetic field in a high-$\sigma$ plasma.

Now let us consider validity of MHD approximation (Eq.(6)) for the
FMS wave. The magnetic field in the pulsar wind is predominantly
toroidal and may be presented as {\bf (from now and hereafter only
averaged quantities will be considered therefore the subscript *
will be omitted)}
$$
B=B_L\frac{R_L}R, \eqno(8)
$$
where $R_L=c/\Omega=5\cdot 10^9 P$ cm is the light cylinder
radius, $P$ the pulsar period. The magnetic field at the light
cylinder may be estimated as
$$
B_L=\frac{\mu}{R_L^3}=9\frac{\mu_{30}}{P^3}\, {\rm G},
$$
where $\mu=10^{30}\mu_{30}$ G$\cdot$cm$^3$ is the magnetic moment
of the star. The wave frequency in the pulsar frame is the pulsar
rotation frequency; in the wind frame the frequency is
$\omega=\Omega/(2\gamma)$. Noting that the magnetic field in the
wind frame is $B'=B/\gamma$, one can reduce the condition (6) to
$$
R<5\cdot 10^{17}\frac{\mu_{30}}P\, {\rm cm}. \eqno(9)
$$
This radius should  be compared with the radius of the standing
shock where the wind terminates. One can place the upper limit on
this radius by equating the magnetic pressure in the wind to the
ram pressure of the medium, $B^2/8\pi=\rho V^2$, where
$\rho=1\rho_0$ g/cm$^3$ is the density of the interstellar gas,
$V=100V_2$ km/s the velocity of pulsar through it. Substituting
Eq.(8), one obtains
$$
R_{\rm shock}=7\times
10^{14}\frac{\mu_{30}}{V_2\sqrt{\rho_0}P^2}\, {\rm cm}\eqno(10)
$$
The radius of the terminating shock may be less than (10) if the
pulsar is surrounded by a dense plerion. Comparing Eq.(10) with
Eq.(9) one can see that FMS waves in the pulsar wind may be
considered in MHD approximation.

The wave may decay in multiple shocks that arise as a result of
nonlinear steepening. Transforming the characteristic nonlinear
scale (7) to the pulsar frame, one gets
$$
R_{\rm nl}=4\gamma^2\sigma R_L.\eqno(11)
$$
The magnetization parameter of the wind depends on the plasma
density. In the pulsar frame, one can conveniently express the
density, $N=n\gamma$, via the dimensionless multiplicity factor,
$\kappa$, as
$$
N=\frac{\kappa\Omega B_L}{2\pi
e}\left(\frac{R_L}R\right)^2.\eqno(12)
$$
Before the multiple shocks arise, the pulsar wind is cold, $w=mn$,
and propagates with a constant Lorentz factor $\gamma_0$. The
magnetization parameter remains constant,
$$
\sigma_0=\frac{B^2}{4\pi
mN\gamma_0}=\frac{\omega_L}{2\kappa\gamma_0\Omega},
$$
where $\omega_L\equiv eB_L/m$ is the gyrofrequency at the light
cylinder. Now one can write the shock formation distance as
$$
R_0=4\gamma_0^2\sigma_0 R_L
=\frac{2\gamma_0\omega_L}{\kappa\Omega^2}=2.5\times
10^{17}\frac{\gamma_0\mu_{30}}{\kappa P}\, {\rm cm} . \eqno(13)
$$
For typical pulsar parameters, $\gamma_0\sim 100$, $\kappa\sim
1-10^3$ (Hibschmann \& Arons 2001), this radius satisfies the
condition (9) therefore MHD description of the wave is justified.
For the typical parameters, this radius exceeds the upper limit on
the termination shock radius (10) for all pulsars with the
exception of the millisecond ones. However there is strong
evidence to suggest that the multiplicity parameter in the Crab
pulsar significantly exceeds the typical one and may be as high as
$\kappa\sim 10^6$ (Shklovsky 1970; Rees \& Gunn 1974). Extra pairs
may be produced together with the observed gamma radiation in the
upper magnetosphere (Cheng, Ho \& Ruderman 1986; Lyubarskii 1996).
At such a plasma density, the multiple shocks arise before the
flow reaches the termination shock (observed in the Crab at $\sim
10^{17}$ cm from the pulsar). So formation of the shocks in the
FMS waves may occur both in the free pulsar wind within the
termination shock and in the pulsar wind nebula beyond the
termination shock.

{\bf If the multiple shocks are not formed until the flow reaches
the termination shock (i.e., if the radius (13) exceeds the
termination shock radius), the waves enter the nebula through the
shock. In case the upstream flow is not Poynting dominated, the
downstream flow is non-relativistic; the nonlinear scale (11) is
extremely small in this case and therefore the waves should decay
immediately, in fact already within the shock. But if the upstream
flow is still Poynting dominated, the shock is weak and the
downstream flow is relativistic (Kundt \& Krotschek 1982, Kennel
\& Coroniti 1984). The waves pass freely the shock from upstream
because reflection is impossible. Parameters of the waves are
nearly not affected by weak shocks (e.g., Anderson 1963) however
the nonlinear scale decreases because the flow decelerates (the
statement by Kennel \& Coroniti that the downstream Lorentz factor
remains large, $\sim\sqrt{\sigma}$, is based on the assumption
that the flow is spherically symmetric; one can see that even
slight deviation of the flow lines from radial may result in
significant deceleration of the flow). Eventually the multiple
shocks are formed. Investigation of the flow in the nebula and
search for the multiple shocks formation zone is out of the scope
of this article. I simply assume that the multiple shocks do form
at some point and consider evolution of the flow parameters in the
course of the wave dissipation. It will be shown in the next
section that in the sub-FMS flow (beyond the termination shock)
the wave energy dissipates immediately after the multiple shocks
arise, dissipation being accompanied by the flow deceleration down
to $\gamma\sim 1$. Hence although the Poynting dominated wind
could not be brought to rest in the termination shock, the wave
dissipation provides an effective mechanism of the flow braking.}

\section{Wave decay}
\subsection{Basic equations}
After the multiple shocks arise, the wave energy dissipates. This
occurs at the nonlinear scale (11), which now may change because
the released energy heats the plasma flow. One can consider
evolution of the flow parameters in the course of the wave decay
applying the conservation laws to the system. Let us separate the
average fluxes of conserving values into the wave and the flow
parts. Such a separation may be performed in general terms for
small amplitude waves (see Appendix 2). The average plasma
density, $n$, and velocity, $v$, (and the corresponding Lorentz
factor $\gamma$) are defined such that the wave does not
contribute to the matter flux (in Appendix 2 the averaged
quantities are marked by tilde; here we work only with averaged
quantities and tilde is omitted). Therefore the continuity
equation may be written as (the flow is considered as locally
spherically symmetric)
$$
n\gamma vR^2=n_0\gamma_0v_0R_0^2, \eqno(14)
$$
where the subscript "0" is referred to the quantities at the shock
formation radius (13). The energy equation may be written as
$$
\frac 1{R^2}\frac
d{dR}\left(w\gamma^2vR^2+\frac{B^2}{4\pi}vR^2\right)=Q, \eqno(15)
$$
where $Q$ is the energy transferred from the wave to unit plasma
volume per unit time. The magnetic field strength may be expressed
via the density making use of the frozen flux condition (see
Eq.(A1.16))
$$
\frac B{Rn\gamma}=\frac{B_0}{R_0n_0\gamma_0}, \eqno(16)
$$
The entropy equation describes the entropy growth as a result of
the wave decay:
$$
T\frac{dS}{d\tau}=p\frac d{d\tau}\frac 1n +3\frac{dT}{d\tau},
\eqno(17)
$$
where $d\tau=dR/\gamma$ is the proper time. We assume that just
after the shock formation, the plasma is heated to relativistic
temperatures, $w=4nT$.

Because the wave propagates in the proper plasma frame with the
velocity $s_M=1-O(1/\sigma)$, its energy and momentum are equal,
in this frame,
 to within $1/\sigma$ (see Appendix 2). Therefore if  the
energy $d\varepsilon$ per unit mass is absorbed in the proper
plasma frame, then the momentum $d\varepsilon$ per unit mass is
also transferred from the wave to the plasma. Then in the pulsar
frame, the absorbed energy is $dE=2\gamma d\varepsilon$. Taking
into account that $dVdt$ is invariant, one can write the energy
release rate as
$$
Q\equiv\frac{dE}{dVdt}=2\gamma n\frac{d\varepsilon}{d\tau}.
$$
Substituting $d\varepsilon=TdS$, one writes the entropy equation
in the form
$$
2\gamma^2\left(3n\frac{dT}{dR}-T\frac{dn}{dR}\right)=Q. \eqno(18)
$$

Let us now turn to the energy equation (15). Making use of
Eqs.(14, 16), one can see that the Poynting flux (the second term
in the brackets in the left-hand side) is nearly independent of
$R$ in the ultrarelativistic case. Expanding this term in
$1/\gamma$ to the next order and retaining in the first term only
leading terms in $1/\gamma$ because this term is already small as
$1/\sigma$, one can reduce Eq.(15) to
$$
\frac
d{dR}\left(w\gamma^2R^2+\frac{B_0^2R_0^2}{8\pi\gamma^2}R^2\right)=QR^2,
\eqno(19)
$$
%In the super-FMS case the second term in the left-hand side is
%small and therefore the wave energy is transmitted to the plasma.
%Taking into account that the wave energy flux is comparable to the
%total Poynting flux, one can see that the plasma energy should
%grow $\sim \sigma$ times.

Substituting $n$ from Eq.(14) into Eqs.(18, 19), one gets finally
$$
\frac
d{dR}\left(\frac{T\gamma}{T_0\gamma_0}+\frac{a\gamma_0^2}{3\gamma^2}
\right)=\frac{QR^2}{4T_0\gamma_0^2n_0R_0^2}; \eqno(20)
$$
$$
3\frac{\gamma}{\gamma_0}\frac{d(T/T_0)}{dR}+ \frac
T{T_0}\frac{d(\gamma/\gamma_0)}{dR}+\frac{2T\gamma}{T_0\gamma_0R}=
\frac{QR^2}{2T_0n_0\gamma_0^2R_0^2}; \eqno(21)
$$
where the factor
$$
a=\frac{3B_0^2}{32\pi
n_0T_0\gamma_0^2}=\frac{3\sigma_0}{2\gamma^2_0}.\eqno(22)
$$
is less than unity in super-FMS flows and exceeds unity in the
opposite case. It was found above that at different wind
parameters, the multiple shocks may arise in FMS waves both before
the wind reaches the termination shock and beyond the shock in the
pulsar wind nebula. Therefore one should consider the wave decay
both in the super-FMS (free wind) and in the sub-FMS (beyond the
termination shock) flows. The flow behavior in these cases is just
opposite therefore let us consider them separately.

\subsection{Wave decay in the super-FMS flow}
The flow in the wind is super-FMS, in this case $a\ll 1$ and the
plasma is cold at $R_0$ so one can take $T_0\sim m$. Eliminating
$Q$ from Eqs.(20, 21) and neglecting the term with $a$, one gets
$$
\frac{\gamma}{\gamma_0}\frac{d(T/T_0)}{dR}- \frac
T{T_0}\frac{d(\gamma/\gamma_0)}{dR}+\frac{2T\gamma}{T_0\gamma_0R}=0,
$$
which yields
$$
\frac{TR^2}{\gamma}=\frac{T_0R_0^2}{\gamma_0}. \eqno(23)
$$
So in the super-FMS flow heating of the medium is accompanied by
acceleration.

The energy release is caused by the nonlinear wave distortion.
Therefore the fraction of the wave energy dissipated at the
distance $\Delta R$ may be roughly estimated as $\Delta R/R_{\rm
nl}$. One should substitute in the expression for the nonlinear
scale (11) parameters varying in the course of the energy
dissipation. Variation of $\sigma$ may be found making use of
Eqs.(14, 16, 23):
$$
\sigma\equiv\frac{B^2}{16\pi
nT\gamma}=\sigma_0\left(\frac{\gamma_0R}{\gamma R_0}\right)^2.
$$
Now the dissipated fraction of the wave energy may be estimated as
$\sim\Delta R/(\sigma\gamma^2 R_L)= R_0\Delta R/R^2$. One can see
that the significant part of the wave energy dissipates at the
scale $\Delta R\sim R_0$. The dissipation scale is roughly equal
to the shock formation scale because acceleration of the flow is
compensated by the plasma heating and corresponding decreasing of
$\sigma$.

Taking into account that the wave energy initially exceeds the
plasma energy about $\sigma_0$ times (for large amplitude waves),
one can see from Eq.(20) (at $a\ll 1$) that
$T\gamma/(T_0\gamma_0)\sim \sigma_0$ after the wave dissipates.
This implies, with the aid of Eq.(23), that $T\sim
m\sqrt{\sigma_0}(R_0/R)$,
$\gamma\sim\sqrt{\sigma_0}\gamma_0(R/R_0)$. So the wave
dissipation at $R\sim R_0$ heats the plasma till the temperature
$\sim m\sqrt{\sigma_0}$ and accelerates it to the Lorentz factor
$\sim\sqrt{\sigma_0}\gamma_0$. Then the heated plasma cools and
accelerates in the Bernoulli regime, $T\propto 1/R$,
$\gamma\propto R$. At the distance $R_1\sim\sqrt{\sigma_0}R_0$,
all the released energy transforms to the kinetic energy and the
plasma reaches the Lorentz factor
$$
\gamma_1\sim\sigma_0\gamma_0=\frac{\omega_L}{2\kappa\Omega}=1.3\times
10^7\frac{\mu_{30}}{\kappa P^2}.
$$

\subsection{Wave decay in the sub-FMS flow}
If the multiple shocks formation radius (13) exceeds the
termination shock radius the waves pass the termination shock and
enter the sub-FMS flow. In the sub-FMS case, the flow decelerates
in the course of energy release (see below) and the nonlinear
scale sharply decreases just after the multiple shocks formation.
Therefore the wave decay scale turns out to be small as compared
with the radius and one can consider the flow in the plane
geometry. In this case Eqs.(20, 21) reduce to
$$
\frac
d{dx}\left(\frac{T\gamma}{T_0\gamma_0}+\frac{a\gamma_0^2}{3\gamma^2}
\right)=1; \eqno(24)
$$
$$
3\frac{\gamma}{\gamma_0}\frac{d(T/T_0)}{dx}+ \frac
T{T_0}\frac{d(\gamma/\gamma_0)}{dx}=2, \eqno(25)
$$
where
$$
x=\int_{r_0}^r\frac Q{4n_0T_0\gamma^2_0}dR
$$
is the ratio of the transferred energy to the initial plasma
energy. When the wave decays, $x$ becomes very large and reaches
the initial ratio of the wave energy to the plasma energy; for
large amplitude waves this ratio is about $\sigma_0$.

Integrating Eq.(24), one yields
$$
\frac{T\gamma}{T_0\gamma_0}=1+\frac a3
+x-\frac{a\gamma_0^2}{3\gamma^2}.
$$
Substituting $T$ from this relation into Eq.(25), one gets the
equation
$$
\frac 2{\gamma}\left(1+\frac
a3+x-\frac{4a\gamma_0^2}{3\gamma^2}\right) \frac{d\gamma}{dx}=1,
$$
which is linear with respect to $x(\gamma)$. The solution is
$$
\frac{\gamma^2}{\gamma_0^2}=\frac{3+a+3x\pm
\sqrt{9(1-a)^2+6x(3+a)+9x^2}}{2(3-a)}, \eqno(26)
$$
where the sign of the square root should be positive for $a<1$
and, correspondingly, negative for $a>1$ to satisfy the initial
conditions. One can see that the super-FMS flow ($a<1$)
accelerates whereas the sub-FMS flow ($a>1$) decelerates in the
course of energy release. For $a>1$, $x\gg 1$ one finds, with the
aid of Eq.(22),
$$
\gamma=\gamma_0\sqrt{\frac{2a}{3x}}=\sqrt{\frac{\sigma_0}{x}};\qquad
T=T_0\sqrt{\frac{3x^3}{8a}}. \eqno(27)
$$
It follows from Eqs.(14, 16) that in the narrow decay zone,
$R-R_0\ll R_0$, the magnetic field remains constant while the flow
remains relativistic, $v\approx 1$. The absorbed energy is
transformed into the internal plasma energy and the magnetization
parameter decreases:
$$
\sigma\equiv\frac{B^2}{4\pi w\gamma^2}=\frac{2\sigma_0}x.
$$
The wave energy dissipates completely when $x\sim\sigma_0\gg 1$.
The nonlinear scale (11) decreases already when $x$ exceeds unity
i.e. when the fraction of the dissipated energy is small ($\sim
1/\sigma_0$) therefore the total dissipation scale turns out to be
small, $\sim R_0/\sigma_0$.

 When the wave decays
completely, $x\sim\sigma_0$ and one gets $\gamma\sim 1$,
$\sigma\sim 1$. However this solution becomes invalid when, in the
proper plasma frame, the particle Larmour radius, $r_g\equiv
T/(eB')$, exceeds the wavelength, $\lambda=2\pi\gamma R_L$. Making
use of Eqs.(8, 12, 22, 27 ), one can write the condition $r_g\sim
2\pi\gamma R_L$ as
$$
\frac x{\sigma_0}\sim\left(\frac{32\pi\kappa
R_L}{R_0}\right)^{2/3}.
$$
The right-hand side of this expression is small therefore
 only a small fraction of the energy is transferred to the plasma in MHD
regime; most of the energy is dissipated when the Larmour radius
exceeds the wavelength. This may result in formation of a high
energy tail in the particle energy distribution.

\section{Conclusions}
In the pulsar wind, a significant fraction of the Poynting flux
may be transported by FMS waves. The plasma in the pulsar wind is
magnetically dominated in the sense that in the proper plasma
frame, the magnetic field energy density exceeds the plasma energy
density. This condition is equivalent to the condition that in the
pulsar frame, the Poynting flux dominates the plasma energy flux.
In such a plasma, FMS waves excite small conductivity current,
oscillations of the magnetic field being nearly compensated by the
displacement current. Thus a very small plasma density is
sufficient to keep this current and  MHD  description of these
waves is justified throughout the pulsar wind till  the
termination shock and beyond.

These waves may decay in multiple shocks that arise through
nonlinear steepening of the waves. Because the conductivity
current is small at the pulsar wind conditions, even a large
amplitude FMS wave is nearly linear and therefore shocks are
formed at large distances from the pulsar. {\bf Depending on the
plasma density in the wind, this may occur either before the flow
reaches the termination shock or beyond the shock. In any case the
wave energy adequately dissipates in the multiple shocks. Energy
release in the super-FMS flow is accompanied by the plasma heating
and acceleration. The thermal pressure also does work on the flow
and the plasma accelerates further on. Therefore if the waves
dissipate in the free, super-FMS pulsar wind, all the wave energy
eventually transforms into the kinetic energy of the flow. If the
waves do not dissipate in the wind, the flow remains Poynting
dominated. In this case the termination shock should be weak and
the downstream flow remains ultra-relativistic. Such a flow does
not match the slow expansion of the nebula; this was considered as
an evidence for conversion of a significant fraction of the
Poynting flux into the particle energy flux before the pulsar wind
reaches the termination shock (Rees \& Gunn 1974; Kennel \&
Coroniti 1984). However it was shown above that the wave
dissipation in the sub-FMS downstream flow  is accompanied by the
flow deceleration to subrelativistic velocities. Hence even if the
flow remains Pointing dominated at the termination shock, the wave
dissipation downstream the shock may provide the necessary
deceleration.}

 Only a fraction of the total Poynting flux is transferred
by the FMS waves therefore the proposed mechanism does not solve
the $\sigma$-problem but should be considered as an element of the
future complete theory. The mechanism considered may also play an
essential role in gamma-ray burst models involving Poynting
dominated outflows from compact objects (Usov 1992, 1994;
M\'esz\'aros \& Rees 1997; Kluzniak \& Ruderman 1997; Blackman \&
Yi 1998; Spruit 1999; Lyutikov \& Blackman 2001; Drenkhahn 2002;
Drenkhahn \& Spruit 2002).

\section*{Acknowledgments}
I am grateful to John Kirk who initiated this research for kind
hospitality during my stay in MPI f\"ur Kernphysik. The work was
supported by MPI f\"ur Kernphysik under their international
visitor program and by Ben Gurion University through a seed grant.

\section*{Appendix 1. Nonlinear FMS waves in a magnetically dominated plasma}

Relativistic nonlinear MHD waves are considered  in the general
case by Akhiezer et al (1975). Here we consider only the waves
propagating perpendicularly to the ambient magnetic field in the
case $\sigma\gg 1$.

Let us first consider the waves in the plane geometry. It is
followed from the continuity equation
$$
\frac{\partial}{\partial t}\gamma n+\frac{\partial}{\partial
x}\gamma nv=0 \eqno(A1.1)
$$
and the frozen-in condition
$$
\frac{\partial B}{\partial t}+\frac{\partial}{\partial}vB=0
$$
that the magnetic field may be presented in the form
$$
B=bn\gamma,\qquad b\equiv\frac{B_*}{n_*\gamma_*}=\it const,
\eqno(A1.2)
$$
{\bf where the subscript * stands for quantities at some fiducial
point; one can conveniently choose this point in such a way that
$B_*$ is equal to $B$ averaged over the wave period.} The dynamic
equation may be presented in the form of the energy equation
$$
\frac{\partial T_{00}}{\partial t}+\frac{\partial T_{01}}{\partial
x}=0, \eqno(A1.3)
$$
where the components of the energy-momentum tensor are
$$
T_{01}=w\gamma^2v+\frac{B^2}{4\pi}\gamma^2 v; \eqno(A1.4)
$$
$$
T_{00}=w\gamma^2 -p+\frac{1+v^2}{8\pi}B^2\gamma^2.\eqno(A1.5)
$$
Note that this set of equations is reduced, with the aid of
Eq.(A1.2), to the standard hydrodynamics equations with
$$
{\cal E}=\varepsilon+\frac{B^2}{8\pi};\quad {\cal
P}=p+\frac{B^2}{8\pi}=Kn^{\Gamma}+\frac{b^2n^2}{8\pi};
$$$$
{\cal W}={\cal E}+{\cal
P}=\varepsilon+p+\frac{B^2}{4\pi}.\eqno(A1.6)
$$

The nonlinear simple wave may be found from the condition that all
dependent variables are functions of one of them, e.g., $n$
(Landau \& Lifshitz 1959). This means that Eq.(A1.3) should be
equivalent to Eq.(A1.1), i.e.
$$
\frac{dT_{01}}{dT_{00}}=\frac{d(n\gamma
v)}{d(n\gamma)}.\eqno(A1.7)
$$
%Introducing the velocity parameter according to the relations
%$$
%v=\tanh\phi;\quad \gamma=\cosh\phi,
%\eqno(A5)
%$$
The last equation reduces to
$$
\gamma^2\frac{dv}{dn}=\frac{s_M}n, \eqno(A1.8)
$$
where
%$$
%s_M^2=\frac{d\cal P}{d\cal E}=\frac{ws^2+b^2n^2/4\pi}
%{w+b^2n^2/4\pi} \eqno(A1.8)
%$$
the FMS velocity $s_M$ is given by Eq.(2). For the small wave
amplitude, $\delta n\equiv n-n_*\ll n_*$, one obtains the linear
FMS wave propagating with the phase velocity $s_M=\it const$. For
a strong wave, $s_M$ depends on the local density therefore the
wave becomes nonlinear. However in the strongly magnetized case,
$\sigma\gg 1$, $s_M$ goes to unity and therefore even a strong
wave is nearly linear.

Substituting $s_M=1$ into Eq.(A1.8), one obtains
$$
\frac n{n_*}=\sqrt{\frac{1+v}{1-v}\frac{1-v_*}{1+v_*}}.
\eqno(A1.9)
$$
In the frame moving, in average, with the plasma, $v_*=0$, one
gets for the small amplitude wave
$$
v=\delta n/n_*.
$$
In the case $\gamma\gg 1$ Eq.(A1.9) reduces to
$$
\gamma=\gamma_* n/n_*.
$$

The waveform moves along the characteristic of Eq.(A1.1):
$$
\frac{dx}{dt}=\frac{d(n\gamma
v)}{d(n\gamma)}=\frac{v+s_M}{1+vs_M}= 1-\frac{2\pi w
n^2_*(1-s^2)}{b^2n^4}\frac{1-v_*}{1+v_*}. \eqno(A1.10)
$$
The last equality in this expression is obtained for the
high-$\sigma$ case. Let us consider the cold plasma, $w=mn$. Then
Eq.(A1.10) yields
$$
x=\left[1-\frac 1{2\sigma_*}\frac{1-v_*}{1+v_*}\left(\frac{
n_*}n\right)^3\right]t+f(n), \eqno(A1.11)
$$
where $\sigma_*\equiv 4\pi m/(b^2 n_*)$. The function $f(n)$ is
determined by the initial waveform. For the initially sinusoidal
wave,
$$
B=B_*\left(1+\alpha\cos\omega x\right),
$$
one gets from Eqs(A1.2, A1.9):
$$
f(n)=\frac{1}{\omega}\arccos\left\{\frac{1+v_*}{2\alpha}\left[
\left(\frac{n}{n_*}\right)^2-1\right]\right\}.\eqno(A1.12)
$$
Note that the dimensionless amplitude, $\alpha$, does not exceed
the value $\alpha_{max}=(1+v_*)/2$; $n\to 0$ at some phase of the
wave period as $\alpha\to\alpha_{max}$.

According to Eq.(A1.10), any portion of the wave advances with a
constant velocity depending only on the plasma density. The
compressive parts overtake the expansive ones and eventually a
shock arises where the waveform becomes vertical,
$$
\frac{dx}{dn}=\frac{d^2x}{dn^2}=0.
$$
This occurs at a time
$$
t_0=\sqrt{\frac{10}3}\frac{2\sigma_*}{9\omega}\frac{1+v_*}{1-v_*}
\frac{\left(4-\sqrt{1+15(\alpha/\alpha_{max})^2}\right)^2}{\sqrt{1+15(\alpha/\alpha_{max})^2}-1}.
\eqno(A1.12)
$$
When the wave amplitude approaches the maximal one, the shock
formation time goes to zero; at small amplitudes, $t_0$ grows as
$1/\alpha$ (Fig.\ 1). {\rm At intermediate amplitudes, the shock
formation time is $t_0\sim\sigma_*/\omega$ in the frame moving
with the plasma, $v_*=0$; if the plasma moves with respect to the
observer with an ultrarelativistic velocity, $\gamma_*\gg 1$, the
shock formation time is $t_0\sim\sigma_*\gamma_*^2/\omega$}.

\begin{figure}
   %includegraphics[width=\textwidth,keepaspectratio]{crab.eps}
\includegraphics[scale=0.4]{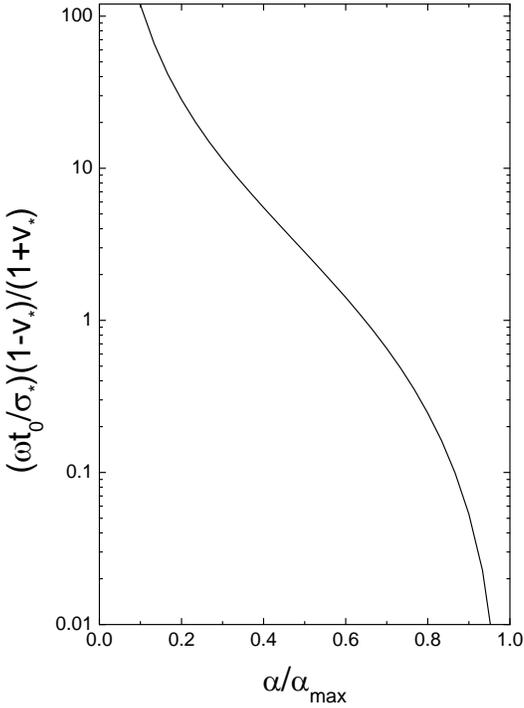}
\caption{Dependence of the shock formation time, $t_0$, on the
wave amplitude, $\alpha$.}
\end{figure}

Now let us consider the radial flow. In this case the governing
equations are written as
$$
\frac{\partial}{\partial t}\gamma n+\frac
1{R^2}\frac{\partial}{\partial R}\gamma R^2nv=0; \eqno(A1.13)
$$
$$
\frac{\partial B}{\partial t}+\frac 1R\frac{\partial}{\partial
R}RvB=0; \eqno(A1.14)
$$
$$
\frac{\partial }{\partial t}T_{00}+\frac 1{R^2}\frac{\partial
}{\partial R}R^2T_{01}=0. \eqno(A1.15)
$$
It follows from Eqs.(A1.13, A1.14) that one can present the
magnetic field as
$$
B=bRn\gamma, \eqno(A1.16)
$$
where $b=\it const$. If the flow is cold, $w=nm$, Eqs.(A1.13,
A1.15, A1.16) reduce to Eqs. (A1.1, A1.2, A1.3) by substitution
$\widetilde n=nR^2$. Therefore the above estimate of the shock
formation time remains valid also for the radial flow.

\section*{Appendix 2. Energy and momentum of the wave}

Let us show that for small amplitude waves, one can separate the
energy and momentum of the wave and the flow without considering
specific waveforms. Let us define the average plasma density, $
n_*$, and velocity, $v_*$,  by the relations
$$
<nv\gamma>=n_* v_*\gamma_*; \eqno(A2.1)
$$
$$
<n\gamma>=n_*\gamma_*, \eqno(A2.2)
$$
where the angular brackets denote averaging over the wave period.
With such a definition, there is no mass flux associated with the
wave.

The total energy and momentum densities are (see Eq.(A1.4-A1.6))
$$
T_{00}={\cal W}\gamma^2-{\cal P}; \eqno(A2.3)
$$
$$
T_{01}={\cal W}v\gamma^2. \eqno(A2.4)
$$
Averaging this values, one can present them as a superposition of
a flow part dependent only on average plasma parameters and a wave
part dependent on the wave amplitude. Calculations are simplified
in the frame where plasma is at rest in average, $v_*=0$.

  In a small amplitude wave, the energy and
momentum are of the second order in the wave amplitude therefore
linear relations, like (see Eq.(A1.8))
$$
\frac{\delta n}{n_*}=\frac{\delta v}{s_M},
$$
may be used only in the second order terms. The average of the
first order terms may be expressed via the second order terms
expanding Eqs.(A2.1, A2.2) in $\delta v$ and $\delta n$ to the
second order. The result is
$$
<\delta n>=-\frac 12 n_*<(\delta v)^2>;
$$
$$
<\delta v>=-\frac 1{s_M}<(\delta v)^2>.
$$
Now expanding Eqs.(A2.3, A2.4) and making use of the
thermodynamical expression $ d{\cal E}/dn={\cal W}/n,$ (because
the wave is isentropic, all thermodynamical values may be
considered as functions of $n$), one yields
$$
<T_{00}>={\cal E}_*+{\cal W}_*<(\delta v)^2>;
$$
$$
<T_{01}>=s_M{\cal W}_*<(\delta v)^2>,
$$
where ${\cal E}_*\equiv {\cal E}(n_*)$ etc. So the energy density
is separated into a part which depends only on the average medium
density and the wave part which is proportional to the wave
amplitude squared. The momentum density of the medium is zero in
the proper plasma frame therefore only the wave momentum
contributes to $<T_{01}>$. The energy and momentum of the wave
coincide, in the proper plasma frame, to within $1-s_M\sim
1/\sigma$. Making the Lorentz transform, one can get components of
the energy-momentum tensor in an arbitrary frame of reference.

\end{document}